\documentclass[a4paper]{jpconf}
\usepackage{graphicx}
\begin{document}
\title{Cluster formation in quantum critical systems}

\author{T. Heitmann$^1$, J. Gaddy$^2$, J. Lamsal$^{1,2}$ and W. Montfrooij$^{1,2}$}

\address{$^1$Missouri Research Reactor, University of Missouri, Columbia, MO 65211 USA\\
$^2$Department of Physics and Astronomy, University of Missouri, Columbia, MO 65211 USA}

\ead{heitmannt@missouri.edu}

\begin{abstract}
The presence of magnetic clusters has been verified in both antiferromagnetic and ferromagnetic quantum critical systems. We review some of the strongest evidence for strongly doped quantum critical systems (Ce(Ru$_{0.24}$Fe$_{0.76}$)$_2$Ge$_2$) and we discuss the implications for the response of the system when cluster formation is combined with finite size effects. In particular, we discuss the change of universality class that is observed close to the order-disorder transition. We detail the conditions under which clustering effects will play a significant role also in the response of stoichiometric systems and their experimental signature.
\end{abstract}

Heavy fermion systems that have been prepared via chemical doping to be at a quantum critical point (QCP) at zero Kelvin display a distribution in Kondo shielding temperatures $T_K$. 
In heavily doped systems, such as UCu$_{5-x}$Pd$_x$ \cite{aronson} and Ce(Ru$_{0.24}$Fe$_{0.76}$)$_2$Ge$_2$ \cite{montfrooij}, the distribution of shielding temperatures can easily span decades of temperature. Upon cooling of such heavily doped systems, the magnetic lattice will be diluted; some magnetic moments will be shielded before others with the result that percolative behavior will ensue. As a consequence, isolated clusters of surviving magnetic ions can and will emerge. Such clusters have indeed been observed in neutron scattering experiments on single crystal Ce(Ru$_{0.24}$Fe$_{0.76}$)$_2$Ge$_2$. Following this direct observation of clusters in this antiferromagnetic (AF) quantum critical compound, the presence of magnetic clusters has since been inferred from uniform susceptibility measurements on various ferromagnetic quantum critical systems\cite{westerkamp, garciasoldevilla}. In fact, the presence of clusters has even been suggested for almost stoichiometric compounds\cite{lausberg}.

In here, we first discuss the importance of finite size effects for isolated magnetic clusters, then we briefly review the evidence for cluster formation in AF Ce(Ru$_{0.24}$Fe$_{0.76}$)$_2$Ge$_2$ and the evidence for finite size effects specifically in this material. Next, we scrutinize finite size effects from both a theoretical and experimental point of view, and we finish by speculating how cluster formation might even shape the response in stoichiometric quantum critical systems.

\section{Finite size effects in isolated magnetic clusters}
\begin{figure}[t]
\begin{minipage}{16pc}
\includegraphics*[viewport=0 -10 450 390,width=70mm,clip]{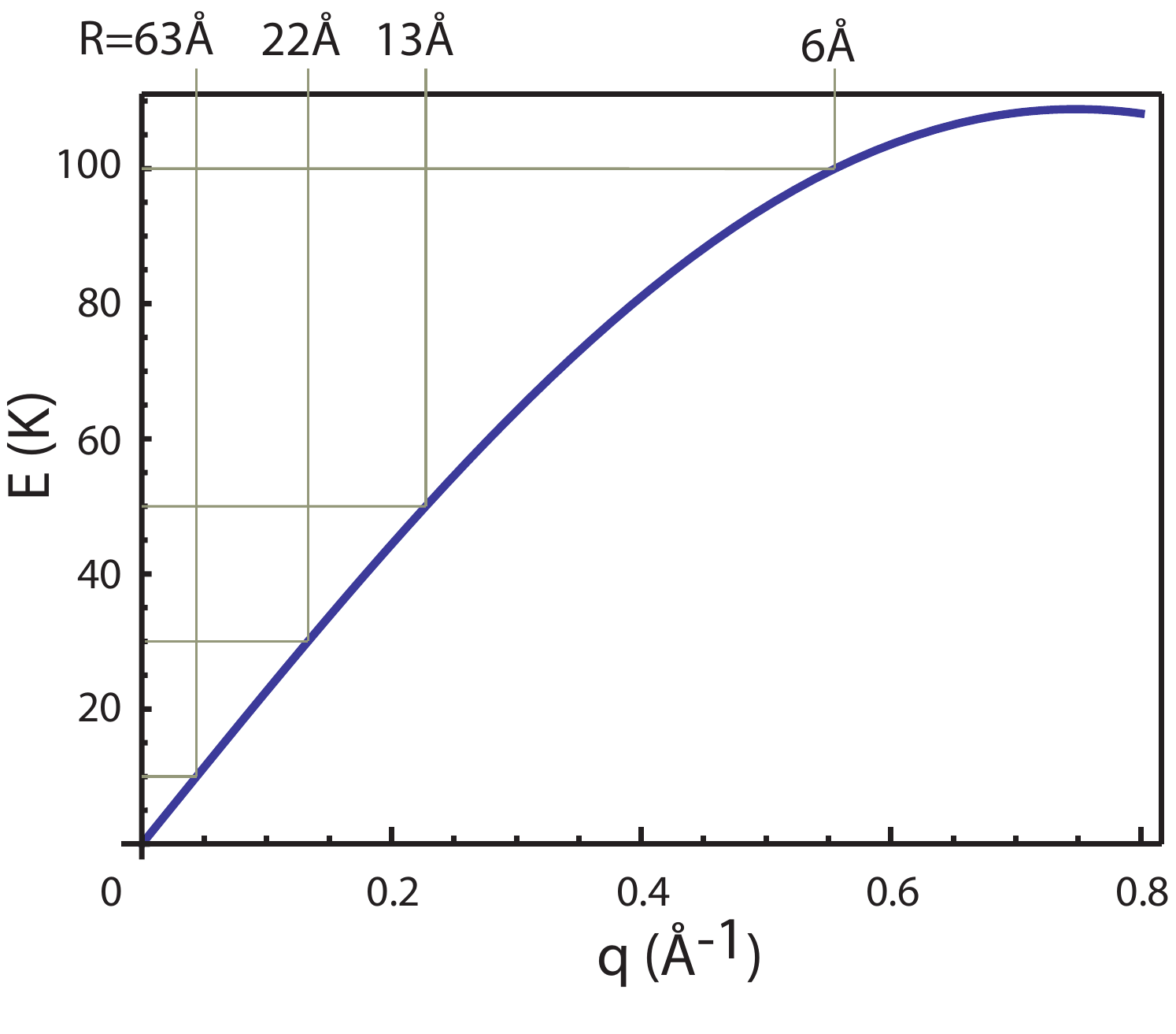}
\caption{\label{dispersion}A typical magnon dispersion for an AF
system, and the  minimum excitation energy induced by finite size
effects. The ordering temperatures of some clusters (with radius $R$
listed at the top of the figure) are shown by the solid lines. For
example, at $T$= 10 K clusters of radius 63$\AA$ and smaller would
order.}
\end{minipage}\hspace{2pc}%
\begin{minipage}{19pc}
\includegraphics*[viewport=0 0 90 80,width=70mm,clip]{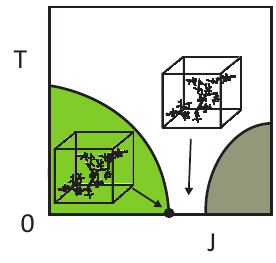}
\caption{\label{phasediagram}The phase diagram for a chemically
doped quantum critical system. Clusters are present both in the
ordered phase (left dome), and in the disordered phase. The right
dome is the region where all moments are shielded. The QCP (black
dot) is the point where a lattice spanning cluster survives down to
$T$= 0 K. (After Ref. \cite{montfrooij})}
\end{minipage}
\end{figure}

The formation of clusters in systems subject to percolation is no surprise, and in heavily doped compounds it is an unavoidable consequence of the dependence of $T_K$ on the immediate surroundings of a magnetic ion. What is perhaps more surprising is that finite size effects (FSE) force all the surviving moments within the isolated cluster to align with their neighbors at relatively modest temperatures. We detail this in the following where we use the dispersion for AF RbMnF$_3$ as an example (Fig. \ref{dispersion}).

Finite size effects come into play when the ordered entity is limited in size, such as is the case for a cluster of radius $R$. The minimum thermal energy required to excite a magnon in a cluster is $k_BT= E(q=2\pi/2R)$. For instance, the minimum magnon energy for a spherical cluster of radius 10\AA (which typically has some 100 members) for RbMnF$_3$ is about 60 K (Fig. \ref{dispersion}). Therefore, at low $T$, we should not expect to find any magnons present in such a cluster and, therefore, the cluster will be ordered. Smaller clusters would undergo this forced ordering at even higher temperatures, clusters of 10,000 members will start to order around $\sim$ 10 K.

This implies that once an isolated cluster materializes, FSE automatically line up the moments within such a cluster. Even when the spin-spin interaction within a cluster has been weakened due to missing (or partially shielded) moments, we can most definitely expect FSE to dominate in the sub-Kelvin range where quantum critical behavior has been studied.


\section{Cluster formation in Ce(Ru$_{0.24}$Fe$_{0.76}$)$_2$Ge$_2$}
The formation of clusters in quantum critical Ce(Ru$_{0.24}$Fe$_{0.76}$)$_2$Ge$_2$ has been observed by means of neutron scattering experiments. Montfrooij et al.\cite{montfrooij} noted the emergence of short range magnetic order upon cooling in experiments on a piece of single crystal that displayed the tell-tale quantum critical response in uniform susceptibility, specific heat and resistivity measurements. The short range order was found to span equal number of moments at all temperatures, independent of crystallographic direction. Since the magnetic body centered Ce-ions are separated by much larger distances along the c-axis than along the a-axis in this tetragonal compound (c/a= 2.5), the only viable explanation for the scattering was that magnetic clusters had formed, and that the moments within these clusters had aligned themselves with their neighbors. This lead the authors to propose a phase diagram that included surviving magnetic moments at $T$= 0 K in the heavy fermion phase (Fig. \ref{phasediagram}).

\section{Percolation theory combined with finite size effects}
The Kondo shielding mechanism effectively removes individual moments from the magnetic lattice at a temperature determined by the local surroundings of the magnetic ion. Invariably this magnetic dilution will lead to the formation of magnetic clusters. Finite size effects, however, do alter the critical behavior in this percolation problem.

The Kondo shielding mechanism is in direct competition with the ordering tendency of the magnetic moments. Once moments have lined up, Kondo shielding will be inhibited by a local molecular field as the shielding mechanism involves spin-flip scattering of the conduction electrons. Therefore, once a cluster becomes isolated from the rest of the magnetic lattice, FSE will line up its moments (Fig. \ref{dispersion}), rendering these clusters impervious to Kondo screening. From a percolation point of view this implies that only moments from the infinite cluster can be removed, isolated clusters cannot be altered.

This restriction on the percolation problem increases the percolation threshold $p_c$ ($p$ is the fraction of occupied sites); every moment removal serves to diminish the strength $P(p)$ of the infinite cluster. It also changes the critical exponents describing the demise of the infinite cluster. This has been calculated in Ref.\cite{gaddy}, but it can easily be reasoned out without resorting to calculations as follows. In standard percolation, moments are removed randomly from any site, including those within finite sized clusters. The closer to the percolation threshold, {\it increasingly} more moments are removed from finite clusters before the infinite clusters loses another moment. On average it will take $p/P$ steps before the infinite cluster is hit again. In contrast, in our restricted scenario it will take only 1 step to remove the next moment. While the infinite clusters end up with an identical morphology in both cases, the difference in occupation probability $p$ at which it occurs {\it increases} by $p/P-1$ at every $p$. Because of the $p/P$ factor, the critical exponents will change. We list the modified critical exponents in Table \ref{critical}.
\begin{table}[t]
\caption{\label{critical}The 3$d$ critical exponents \cite{gaddy} for standard and restricted percolation. The last column shows the exact scaling relationships; primed values reflect the restricted percolation case.}
\begin{center}
  \begin{tabular}{l l l l}
    \hline
    &standard& restricted & relationship \\[1pt]

      \hline
      \\[-6pt]
   $\alpha$& -0.62 &1& analytic \\[3pt]
    $\beta$& 0.41 & 0.205& $\beta'=\beta/2$ \\[3pt]
    $\gamma$& 1.80 &0.595& $\gamma'=1-\beta$\\[3pt]
     $\nu$& 0.88 & 1/3 & $\nu'=1/d$ \\ 
 \hline
  \end{tabular}
  \end{center}
\end{table}


Given that a distribution of $T_K$ must necessarily lead to a restricted percolation scenario describing the dilution of the magnetic lattice near a QCP, we expect to see the critical exponents listed in Table \ref{critical} to be reflected in the response of the system. The exact reflection will depend on the distribution of $T_K$: the exponents in Table \ref{critical} are given as a function of $p$, whereas in experiments they are measured as a function of $T$. The Kondo distribution determines $p(T)$, and the low temperature part of $p(T)$-well below the average Kondo shielding temperature- will determine the critical exponents as a function of temperature\cite{stauffer}. The details have not been resolved yet, however, the overall mechanism of how the critical behavior near the QCP is determined by the emergence of magnetic clusters has now been qualitatively established.

\section{Clusters in stoichiometric systems}
We argue that it is much more likely than not that magnetic clusters also shape the response in (nearly) stoichiometric quantum critical systems. Close to any phase transition, fluctuations of the order parameter will result in ordered regions. Close to a QCP, we can expect fluctuations in the tendency of individual moments to become shielded. These fluctuations can happen because of the bosonic degrees of freedom (zero-point-motion changing the interatomic separation with a concomitant change in $T_K$) or because of the fermionic degrees of freedom (associated with the shielding conduction electrons themselves). Because of such fluctuations, regions materialize that have magnetic moments isolated from the rest of the lattice by shielded moments; close to the phase transition, such regions (clusters) can contain large numbers of moments.

Once large clusters form, finite size effects dictate that the moments within these clusters order. The main difference with heavily doped systems is that these clusters have a limited lifetime; this lifetime determines whether these clusters will shape the overall response of the system. Typically, zero-point-motion induced atomic displacements happen on a much slower time scale than the electronic motion; in this case, clusters will shape the electronic response of the system. For the case of clusters finding their origin in the fermionic degrees of freedom we cannot make such a statement. However, it is clear that clusters should be expected to emerge, what is unclear is their exact influence on the response of the system.

Emerging clusters in stoichiometric systems can be stabilized with the aid of an external magnetic field. When isolated clusters form, they will likely have dangling moments, bestowing the entire cluster with a net magnetic moment even in the case of AF-coupling. An external field lowers the energy of such clusters to the point that they can last long enough to dominate the response of the system. A good candidate for such a field stabilized system would be YbRh$_2$Si$_2$\cite{custers}. We finish with a list of experimental signatures that would be observed should clusters indeed form in stoichiometric systems.

First, neutron scattering experiments would observe the emergence of
short range order characterized by equal numbers of magnetic moments
coupled along any crystallographic direction, independent of actual
lattice spacings. Second, once clusters form, we should see a sharp
drop in resistivity since conduction electrons will no longer be
scattered through the Kondo mechanism in these clusters. Third, we
expect to see the formation of the clusters reflected in the
specific heat given the loss of entropy once the moments within
clusters line up. Fourth, the uniform susceptibility will reflect
the presence of field stabilized magnetic clusters, and a strong
signal will be expected given the net moment of most clusters. The
response should also show a ferromagnetic component, once again
reflecting dangling moments of fully aligned clusters. All
measurements should reflect the underlying percolation critical
exponents.

Standard tests for the presence of magnetic clusters might not work
in quantum critical systems. The clusters do not live forever, and
hence, standard hysteresis measurements might only see clusters
appear, disappear and appear again without any hysteresis.
Resistivity measurements are expected to show a sharp drop anyways
once the coherence temperature is reached. The clusters are not
expected to provide a measurable residual resistivity as there is no
real disorder in the system; rather, the clusters provide low
resistance pathways for the conduction electrons to go through, or
even to diffract around. Neutron scattering experiments will provide
unambiguous evidence, or lack thereof, for the formation of
clusters. For now, it is interesting to note that the observed
exponents for the specific heat and susceptibility in
YbRh$_2$Si$_2$\cite{custers} exactly coincide with the ones given in
Table \ref{critical} and calculated in Ref \cite{gaddy}.\\

This research is supported by the U.S. Department of Energy, Basic
Energy Sciences, and the Division of Materials Sciences and
Engineering under Grant No.DE-FG02-07ER46381.

\end{document}